\documentclass[apj]{emulateapj}

\def\gsim{\;\rlap{\lower 2.5pt
 \hbox{$\sim$}}\raise 1.5pt\hbox{$>$}\;}
\def\lsim{\;\rlap{\lower 2.5pt
   \hbox{$\sim$}}\raise 1.5pt\hbox{$<$}\;}
\def\ie{{\it i.e. }}
\def\eg{{\it e.g. }}

\shorttitle{Testing Shear Recovery with Field Distortion}
\shortauthors{Zhang et al.}

\begin{document}
\title{Testing Shear Recovery with Field Distortion}

\author{Jun Zhang$^{1,2*}$, Fuyu Dong$^1$, Hekun Li$^1$, Xiangchong Li$^{3,4}$, Yingke Li$^1$, Dezi Liu$^{5,6,7}$,\\ Wentao Luo$^1$, Liping Fu$^6$, Guoliang Li$^{8,9}$, Zuhui Fan$^{5,7}$}
\affil{
$^1$Department of Astronomy, Shanghai Jiao Tong University, Shanghai 200240, China \\
$^2$Shanghai Key Laboratory for Particle Physics and Cosmology, Shanghai 200240, China \\
$^3$Department of Physics, University of Tokyo, Tokyo, 113-0033, Japan\\
$^4$Kavli Institute for the Physics and Mathematics of the Universe (Kavli IPMU), UTIAS,\\ Tokyo Institutes for Advanced Study, University of Tokyo, Chiba, 277-8583, Japan\\
$^5$South-Western Institute for Astronomy Research, Yunnan University, Kunming 650500, China\\
$^6$The Shanghai Key Lab for Astrophysics, Shanghai Normal University, Shanghai  200234, China\\
$^7$Department of Astronomy, School of Physics, Peking University, Beijing 100871, China\\
$^8$Purple Mountain Observatory, Chinese Academy of Sciences, Nanjing 210000, China\\
$^9$School of Astronomy and Space Sciences, University of Science and Technology of China, Hefei, 230029, China
}

\email{*betajzhang@sjtu.edu.cn}

\begin{abstract}

The tilt, rotation, or offset of each CCD with respect to the focal plane, as well as the distortion of the focal plane itself, cause shape distortions to the observed objects, an effect typically known as field distortion (FD). We point out that FD provides a unique way of quantifying the accuracy of cosmic shear measurement. The idea is to stack the shear estimators from galaxies that share similar FD-induced shape distortions. Given that the latter can be calculated with parameters from astrometric calibrations, the accuracy of the shear estimator can be directly tested on real images. It provides a way to calibrate the multiplicative and additive shear recovery biases within the scientific data itself, without requiring simulations or any external data sets. We use the CFHTLenS images to test the Fourier\_Quad shear recovery method. We highlight some details in our image processing pipeline, including background removal, source identification and deblending, astrometric calibration, star selection for PSF reconstruction, noise reduction, etc.. We show that in the shear ranges of $-0.005\lsim g_1\lsim 0.005$ and $-0.008\lsim g_2\lsim 0.008$, the multiplicative biases are at the level of $\lsim 0.04$. Slight additive biases on the order of $\sim 5\times 10^{-4}$ ($6 \sigma$) are identified for sources provided by the official CFHTLenS catalog (not using its shear catalog), but are minor ($4 \sigma$) for source catalog generated by our Fourier\_Quad pipeline.  
\end{abstract}


\section{Introduction}
\label{intro}

Recent weak lensing measurements have indicated marginal tensions with results from the cosmic microwave background \citep{hildebrandt2017,kohlinger2017,DES1,DES2,HSC1}, and galaxy formation models \citep{leauthaud2017}. It would be extremely exciting if these discrepancies point to new physics beyond the concordance $\Lambda$CDM cosmological model, e.g., dynamical dark energy or modified gravity theories \citep{amendola2018}. There are, however, also concerns that weak lensing measurements still contain systematic errors that have not been identified and corrected, regarding either shape measurement or photo-z errors \citep{EL2018}.

To establish weak lensing as a robust cosmological probe, ways of testing the accuracy and consistency of the measurements are indispensable. Simulations are typically used to calibrate the shear recovery accuracy \citep{great3}, but image processing at different stages could all introduce problems that are not easily identified, and therefore not included in the simulations. 

There are also ways of testing the consistency of shear recovery based on real data. For example: cross-correlation between the PSF shape and the galaxy shape is used to find out if there are residual anisotropic PSF effects that are not removed \citep{kaiser1995,fischer2000}; E/B mode decomposition is used to check if the lensing signal has a gravitational origin \citep{crittenden2002,schneider2002}. Nevertheless, we find that these tests are not very sensitive to the multiplicative biases that people typically require to correct for the sensitivity of their shear estimator to the underlying shear signal. Given that the multiplicative biases are directly degenerate with the amplitudes of the lensing signals and the shear-shear correlations, and thereby several key cosmological parameters such as $\sigma_8$ and $\Omega_m$, it is desirable to calibrate the multiplicative biases based on real data as an alternative to the existing methods \citep{vallinotto2011,zhangpengjie2015}. The purpose of this work is to propose a solution with the help of field distortion, an effect that exists universally in optical systems. 

The effect of field distortion modifies the galaxy image in a way similar to that of lensing. The amplitude and direction of this distortion can be calculated using the parameters derived from astrometric calibration on one hand, and recovered from the galaxy images on the other hand. We therefore have a natural way of testing the accuracy of shear recovery directly using real data, without requiring simulations or external data sets. 

In \S\ref{fd}, we introduce the concept of field distortion, and the way of deriving it using the astrometric parameters. In \S\ref{shear_measure}, we give a brief introduction of the Fourier\_Quad shear measurement method \citep{zlf2015}. We use the CFHTLenS data \citep{heymans2012,erben2013} to demonstrate our idea. The details regarding our image processing of the CFHTLenS data are given in \S\ref{reduction}. It highlights a few key steps in our pipeline. Our main results are shown in \S\ref{results}. Finally, in \S\ref{conclusion}, we give brief conclusions as well as discussions regarding the current status of our pipeline and future prospects.

\section{Field Distortion}
\label{fd}

In geometric optics, field distortion is a type of optical aberration. In general, it is a result of deviation from global rectilinear projection, causing distortions of image shapes. Here in this paper, we are interested in the distortion at any local position of the CCD, i.e., the vicinity of a galaxy. The local distortion can always be treated as a linear mapping between the source plane and the CCD plane. If one thinks of this effect in terms of the displacements of point sources, it is straightforward to draw an analogy between the field distortion and the weak lensing effect. In other words, field distortion causes a shear signal on top of the astrophysical one. It is our purpose to utilize this signal to test the accuracy of shear measurement.

In this work, we adopt the TPV world coordinate system (WCS) \citep{cg2002}, which builds on the standard tangent plane projection with a general polynomial distortion correction. This is the convention used in the CFHTLenS data processing. Our notations below can be easily extended to other WCS conventions. Let us denote the standard coordinate of the source on the tangent plane of the celestial sphere as $(\xi,\eta)$, and its position on the CCD grid as $(x,y)$. They are related through the following formulae in the TPV convention:
\begin{eqnarray}
\label{XYxy}
&&\xi=f_{\xi}(X,Y) \\ \nonumber
&&\eta=f_{\eta}(X,Y)\\ \nonumber
&&X(x,y)=\mathrm{CD^1_1}(x-{\mathrm{CRPIX1}})+\mathrm{CD^1_2}(y-\mathrm{CRPIX2}) \\ \nonumber
&&Y(x,y)=\mathrm{CD^2_1}(x-{\mathrm{CRPIX1}})+\mathrm{CD^2_2}(y-\mathrm{CRPIX2})
\end{eqnarray}
In which the coefficients $\mathrm{CD^i_j}$ and $\mathrm{CRPIX(1,2)}$ take care of the basic linear transformation, and the distortion functions $f_{\xi}$ and $f_{\eta}$ are defined in polynomial forms as:
\begin{eqnarray}
\label{mapping}
&&f_{\xi}(X,Y)\\ \nonumber
&=&\mathrm{PV^1_0}+\mathrm{PV^1_1}X+\mathrm{PV^1_2}Y+\mathrm{PV^1_3}R\\ \nonumber
&+&\mathrm{PV^1_4}X^2+\mathrm{PV^1_5}XY+\mathrm{PV^1_6}Y^2+\mathrm{PV^1_7}X^3\\ \nonumber
&+&\mathrm{PV^1_8}X^2Y+\mathrm{PV^1_9}XY^2+\mathrm{PV^1_{10}}Y^3+\mathrm{PV^1_{11}}R^3+... \\ \nonumber
&&f_{\eta}(X,Y)\\ \nonumber
&=&\mathrm{PV^2_0}+\mathrm{PV^2_1}Y+\mathrm{PV^2_2}X+\mathrm{PV^2_3}R\\ \nonumber
&+&\mathrm{PV^2_4}Y^2+\mathrm{PV^2_5}XY+\mathrm{PV^2_6}X^2+\mathrm{PV^2_7}Y^3\\ \nonumber
&+&\mathrm{PV^2_8}XY^2+\mathrm{PV^2_9}X^2Y+\mathrm{PV^2_{10}}X^3+\mathrm{PV^2_{11}}R^3+... 
\end{eqnarray}
with $R=\sqrt{X^2+Y^2}$. Eq.(\ref{mapping}) can in principle include many more higher order polynomial terms. The CFHTLenS team includes only the coefficients of the form $\mathrm{PV^{1,2}_i}$, with $i=0,1,2,4,5,6,7,8,9,10$.

\begin{figure}
  \centering
  \includegraphics[width=8cm]{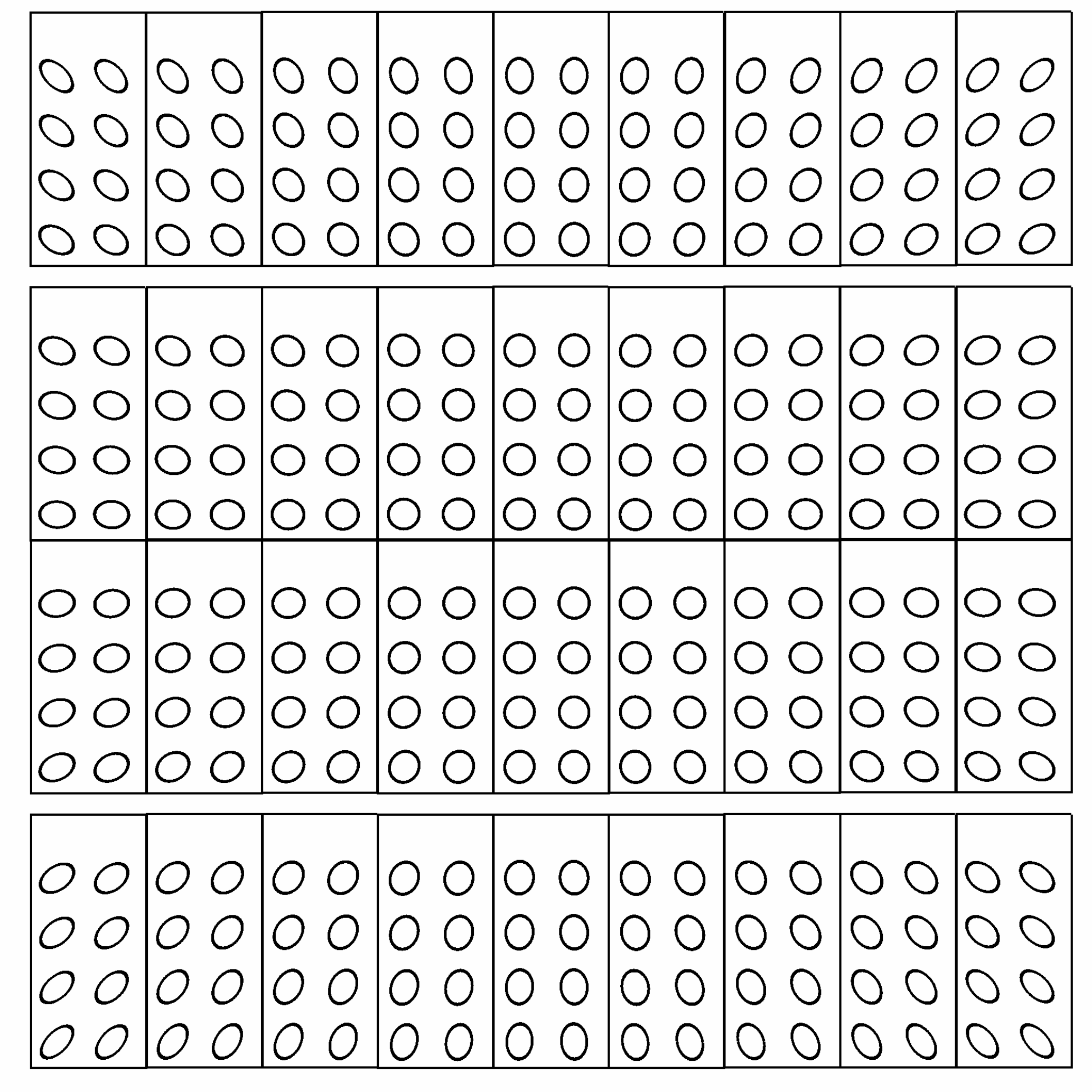}
  \caption{The distribution of the FDS signals on the $4\times 9$ CCD's of one MegaCam exposure. The ellipticities are enlarged by 30 times for a better visibility. }
  \label{FD}
\end{figure}

One can directly derive the shape distortion matrix from the above formulae. As in weak lensing, the matrix can be written as:
\begin{eqnarray}
\left(\begin{array}{c}
d\xi \\
d\eta \end{array}\right)
=\left(\begin{array}{cc}
\partial_x \xi &\;\; \partial_y\xi \\
\partial_x \eta &\;\; \partial_y\eta \end{array}\right)
\left(\begin{array}{c}
dx \\
dy \end{array}\right)
\end{eqnarray}
The distortion matrix can usually be decomposed into four parts: an overall scaling factor $M$ (or magnification), two shear components $(g_1, g_2)$, and a rotation angle $\theta$. We can rewrite the distortion matrix in the following form ($g_1, g_2, \theta \ll 1$):
\begin{eqnarray}
\label{decompose2}
\left(\begin{array}{cc}
\partial_x \xi &\;\; \partial_y\xi \\
\partial_x \eta &\;\; \partial_y\eta \end{array}\right)
=M\left(\begin{array}{cc}
1-g_1 &\;\; -g_2+\theta \\
-g_2-\theta &\;\; 1+g_1 \end{array}\right)
\end{eqnarray}
It is then straightforward to derive the Field-Distortion-induced Shear (called FDS hereafter) as:
\begin{eqnarray}
&&g_1(\mathrm{FD})=(\partial_y\eta-\partial_x\xi)/(\partial_y\eta+\partial_x\xi)\\ \nonumber
&&g_2(\mathrm{FD})=-(\partial_x\eta+\partial_y\xi)/(\partial_y\eta+\partial_x\xi)
\end{eqnarray}
Fig.\ref{FD} shows a typical distribution of the FDS on the 36 CCD's of MegaCam at CFHT\footnote{http://www.cfht.hawaii.edu/Instruments/Imaging/Megacam/}. For the purpose of illustration, the FDS at each location is magnified by a factor of 30, and represented with an ellipse. The parameters of the field distortion are derived from astrometric calibration done with the THELI software \citep{erben2005,schirmer2013} using the reference catalog from either 2MASS or SDSS. The amplitude of the FDS here is typically a few times $0.001$, similar to the weak lensing signal. The purpose of this work is to find out if one can recover the FDS signals with the galaxy images, as a way of checking the accuracy of the shear measurement method. Note that as the FDS only depends on the location of the galaxy on the CCD, the cosmological weak lensing signal can thus be regarded as a random noise, similar to the galaxy intrinsic shape variation.   

It is worth noting that in the Lensfit pipeline of CFHTLenS \citep{miller2007,miller2013}, the field distortion signal is added to the assumed model of the galaxy on each exposure before performing the joint fitting for the galaxy parameters/ellipticities. This is for the purpose of increasing the significance of the best-fitting model, thereby reducing the errors on the galaxy ellipticities. However, in doing so, the FDS information is lost. In this work, we use the shear measurement method developed through a series of work \citep{zhang2008,zhang2010,zhang2011,zk2011,zlf2015,zzl2017}. It is named ``Fourier\_Quad'' in the GREAT3 project \citep{great3}. Our shear measurements are performed on individual exposures, therefore the FDS information can be completely retained. In the rest of the paper, we introduce the Fourier\_Quad method, and the test of it with FDS using the CFHTLenS images.

\section{The Fourier\_Quad Method}
\label{shear_measure}

In Fourier\_Quad, the shear estimators are defined on the 2D power spectrum of the galaxy image in Fourier space:
\begin{eqnarray}
\label{shear_estimator}
G_1&=&-\frac{1}{2}\int d^2\vec{k}(k_x^2-k_y^2)T(\vec{k})M(\vec{k})\\ \nonumber
G_2&=&-\int d^2\vec{k}k_xk_yT(\vec{k})M(\vec{k})\\ \nonumber
N&=&\int d^2\vec{k}\left[k^2-\frac{\beta^2}{2}k^4\right]T(\vec{k})M(\vec{k})
\end{eqnarray}
where $\vec{k}$ is the wave vector. $T(\vec{k})$ is given by $\left\vert\widetilde{W}_{\beta}(\vec{k})\right\vert^2/\left\vert\widetilde{W}_{PSF}(\vec{k})\right\vert^2$, \ie, the ratio between the power spectrum of a 2D isotropic Gaussian function $W_{\beta}$ and that of the point spread function (PSF) $W_{PSF}$. This factor is used to convert the form of the PSF to the isotropic Gaussian function for the purpose of correcting the PSF effect in a model-independent way. The Gaussian function $W_{\beta}$ is defined as:
\begin{equation}
W_{\beta}(\vec{x})=\frac{1}{2\pi\beta^2}\exp\left(-\frac{\left\vert\vec{x}\right\vert^2}{2\beta^2}\right).
\end{equation} 
in which $\beta$ should be somewhat larger than the scale radius of the original PSF to avoid singularities in the conversion. $M(\vec{k})$ is the galaxy power spectrum, but modified to take into account the corrections due to the background and the Poisson noise:
\begin{eqnarray}
\label{TM}
&&M(\vec{k})=\left\vert\widetilde{f}^S(\vec{k})\right\vert^2-F^S-\left\vert\widetilde{f}^B(\vec{k})\right\vert^2+F^B\\ \nonumber
&&F^S=\frac{\int_{\vert\vec{k}\vert > k_c} d^2\vec{k}\left\vert\widetilde{f}^S(\vec{k})\right\vert^2}{\int_{\vert\vec{k}\vert > k_c} d^2\vec{k}}, \;\;\; F^B=\frac{\int_{\vert\vec{k}\vert > k_c} d^2\vec{k}\left\vert\widetilde{f}^B(\vec{k})\right\vert^2}{\int_{\vert\vec{k}\vert > k_c} d^2\vec{k}}
\end{eqnarray}
$\widetilde{f}^S(\vec{k})$ and $\widetilde{f}^B(\vec{k})$ are the Fourier transformations of the galaxy image and a neighboring image of background noise respectively. The terms $F^S$ and $F^B$ serve as good estimates of the Poisson noise power spectra on the source and background images respectively, given that the critical wave number $k_c$ is large enough to avoid the regions dominated by the source power. It has been shown in \cite{zlf2015} that the ensemble averages of the shear estimators defined above recover the shear values to the second order in accuracy (assuming that the intrinsic galaxy images are statistically isotropic), \ie, 
\begin{equation}
\label{shear_measurement}
\frac{\left\langle  G_1\right\rangle }{\left\langle  N\right\rangle }=g_1+O(g_{1,2}^3),\;\;\;\frac{\left\langle  G_2\right\rangle }{\left\langle  N\right\rangle }=g_2+O(g_{1,2}^3)
\end{equation}
Note that the ensemble averages are taken for $G_1$, $G_2$, and $N$ separately \citep{zk2011}. 

A more refined way of deriving the shear signal from an ensemble of shear estimators is given in \cite{zzl2017}. The method is called PDF-SYM, in which the best estimate of the shear signal $(\hat{g}_1, \hat{g}_2)$ is determined by symmetrizing the probability distribution function (PDF) of $G_1-\hat{g}_1(N+U)$ and $G_2-\hat{g}_2(N-U)$ on the positive and negative sides of zero. For this purpose, two more terms should be defined to properly take into account the parity properties of the shear estimators:
\begin{eqnarray} 
\label{def_U} 
U&=&-\frac{\beta^2}{2}\int d^2\vec{k}\left(k_x^4-6k_x^2k_y^2+k_y^4\right)T(\vec{k})M(\vec{k})\\ \nonumber
V&=&-2\beta^2\int d^2\vec{k}\left(k_x^3k_y-k_xk_y^3\right)T(\vec{k})M(\vec{k})
\end{eqnarray}
The term $V$ is kept for transforming $U$ in case of coordinate rotation in shear measurement. It turns out that the PDF-SYM method allows the statistical error of the shear measurement to approach the Cramer-Rao bound, \ie, the lower limit in theory. We adopt the PDF-SYM method for presenting the shear measurement results in this paper. 

Note that in astrophysical applications, we need to remove the field-distortion effect from the shear estimators. For this purpose, we only need to replace $(G_1,G_2)$ with $(G_1',G_2')$ that are defined as:
\begin{eqnarray}
&&G_1'=G_1-g_1(\mathrm{FD})(N+U)-g_2(\mathrm{FD})V\\ \nonumber
&&G_2'=G_2-g_2(\mathrm{FD})(N-U)-g_1(\mathrm{FD})V
\end{eqnarray}

\section{Image Processing with the CFHTLenS Data}
\label{reduction}

Our imaging data is from the Wide part of CFHTLS observed in 2003 and 2008 with MegaPrime/MegaCam \citep{erben2013}. MegaCam is an optical multi-chip instrument with a $9\times 4$ CCD array, $0.187'' $ pixel scale, and $\sim 1^{\circ}\times 1^{\circ}$ field-of-view. There are four fields (W1,2,3,4) in the survey, containing 171 pointings in total, covering about 154 $\mathrm{deg^2}$ sky area. There are imaging data for five filters: u*,g',r',i',z', and we use the i'-band data for shear measurement in this paper\footnote{We only exclude bad exposures in our analysis, including exposure ``827410'' of w1m2m2, ``792617'' of w3m1m0, ``792436'' of w3m1m2, ``987104'' of w3p2m3, ``859948'' of w4m1p0, ``859950'' of w4m1p0, and the whole field of w2p2p2.}. For the i'-band, each pointing typically contains seven exposures, each of which lasts for 615 seconds. The limiting magnitude in the i'-band reaches around 24.5 (AB mag). The images we use are all preprocessed with the Elixir software\footnote{http://www.cfht.hawaii.edu/Instruments/Elixir/}, which takes care of the removal of instrumental signatures \citep{mc2004}. These images are available at the Canadian Astronomical Data Center (CADC)\footnote{http://www4.cadc-ccda.hia-iha.nrc-cnrc.gc.ca/cadc/}. 

Unlike the Lensfit method used by the CFHTLenS team, our Fourier\_Quad method is so far carried out on single exposures individually. Initially in this project, we rely on the THELI software to process the CCD images, mainly regarding background removal, defect detection, and astrometric calibration. We skip the steps of photometric calibration and image co-addition. As our project evolves, we feel more and more obligatory to develop a self-contained image processing pipeline, including all the necessary steps for ensuring shear recovery accuracy, so that sources of shear biases from individual image processing steps can potentially be targeted and corrected. It is worth noting that several crucial problems are indeed identified with the field-distortion test proposed in this paper. Our results in the rest of this paper are generated using our newly-built image processing pipeline, which is completely independent from the THELI software or any other astronomical softwares such as, \eg, Sextractor \citep{ba1996}. It works directly on the Elixir-processed images, and does background removal, defects-identification, source identification, deblending of sources, astrometric calibration, PSF reconstruction, and shear measurement. Our code does not yet include photometric calibration, therefore it does not measure magnitude or photometric redshift. As a remedy for this, our code can run shear measurements for sources whose positions in terms of (RA, DEC) are from an external source catalog. Some of our results are made with the official CFHTLenS source catalog, as shown latter in the paper. A complete description of our pipeline will be given in a separate paper (Li et al., in preparation). In the following sections, we show several key highlights.

\subsection{Background Removal}

Accurate modeling of the background is important not only for source identification, but also for shear measurement. When the discrete sources are not over-crowded, which is the case for the individual exposures of CFHTLenS, a common way to find the local background level is by sorting a reasonable number of neighboring pixel values, followed by removing the outliers that are either due to bright sources or bad pixels, and finally equating the background as a linear combination of the median and the mean of the remaining pixels \citep{ba1996}. The chip-wide background map is constructed with a bicubic-spline interpolation of a number of local background values, each of which is derived from a region of a reasonable mesh size. We adopt a similar approach in our pipeline for making the background map, but with a polynomial fitting for the chip-wide map. In doing so, we remove the outliers of the background values that are typically associated with extended diffraction patterns of bright stars, and repeat the fitting until there are no more outliers. The distribution of the standard deviation $\sigma(x,y)$ of the background is similarly generated.  

The background-subtracted CCD images created with the above procedures generally perform well for source identification. Nevertheless, we find that removal of the chip-wide background map inevitably leaves spatially-varying residuals, causing excess anisotropies in the source stamps and shear biases. In order to avoid this problem, we choose to cut out the source stamps directly from the original CCD image. For each source stamp ($48\times 48$), we locally model its background as a linear 2D function, with parameters evaluated using the pixels on the immediate neighborhood of the source stamp. We exclude the bad pixels and the bright pixels from other sources by sorting the pixel values. It turns out that source stamps created in this way perform very well in shear measurement.

It is useful to note that each of the 36 CCD's of Megacam is a combination of two equal-sized chips lining up along the short axis. Their gains sometimes have visible differences due to imperfect flat-fielding process by Elixir. For avoiding systematic shear errors caused by this feature of the data, we simply remove sources that span both sides of a CCD.

\subsection{Source Identification and Deblending}
\label{source_select}

There are two stages in our source identification: 1. locating the source positions in the CCD; 2. determining the boundaries of the sources. If an external source catalog is provided, the first step is skipped. Otherwise, each source is identified as a number of connected pixels that are above $n\sigma$ using the FOF (Friends-Of-Friends) method. We also require every source FOF region to contain at least one pixel that is above $m\sigma$ ($m>n$), and the number of pixels in each group should be larger than $s$. In the work of this paper, we choose $n=2, m=4, s=6$. Note that unlike in Sextractor \citep{ba1996}, we try not to involve a filter of a certain size (\eg, $3\times 3$, $5\times 5$) and shape in source finding, for the purpose of maximally avoiding potential selection effects. For avoiding possible nonlinear-response regions of the CCD, we exclude sources with pixels brighter than $1/2$ of the saturation level.  

\begin{figure}
  \centering
  \includegraphics[width=8cm]{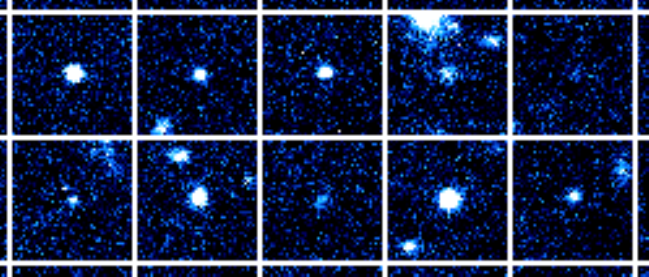} \\ \vspace{10pt}
  \includegraphics[width=8cm]{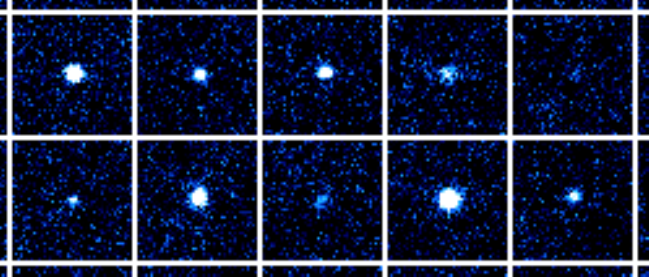}
  \caption{Examples showing how our pipeline removes contamination from neighboring sources. The upper panel shows the original appearance of stamps cut out from the CCD image, and the lower panel shows the same stamps after the removal of the neighboring contaminants to the central sources. }
  \label{blend}
\end{figure}

In the second step, each source is required to be well contained within a $48\times 48$ stamp, meaning that the source pixels should stay inside a circular region of radius $24-a$ (pixel) centered at the stamp center. $a$ is typically less than $5$. For other source regions or masked areas inside the stamp, we replace their pixel values with uncorrelated Gaussian random noises of variance $\sigma^2(x,y)$, where $(x,y)$ refers to the central position of the stamp. If the masked area directly touches the region of the target source, we simply consider it as a bad image, and remove the source from the catalog. In fig.\ref{blend}, we show the repaired source stamps as an example. 

In this work, blending effectively means that the FOF regions of two sources are overlapped. In Fourier\_Quad, deblending is not necessary if the two blended sources are physically close, \ie, their redshifts are similar. This is because the accuracy of the Fourier\_Quad method does not rely on the regularity of the source shape. If the blended sources have very different redshifts, we need to carefully interpret the shear results based on the frequency of this case, or simply remove them from the source list. However, it is not yet clear how accurately the photometric redshifts for two or more blended sources can be measured individually in practice. This is by itself a difficult problem, and beyond the scope of this work. For the purpose of this paper, we simply treat each FOF group as a single source, as it does not affect the recovery of the FDS.   
 
Finally, according to the method of Fourier\_Quad, we need to find a nearby stamp of background noise for each source image. Currently, our choice is to locate several candidate stamps in the source neighborhood, and pick the one whose maximal pixel value is the lowest. Once the noise stamp is chosen, any source pixels and bad pixels identified within it are replaced with randomly generated noise pixels, similar to what is done for the source stamp. Note that the undetected sources/blends are effectively treated as a part of the background. As long as their contributions to the power spectra of the source stamps and the noise stamps are statistically similar, they should not cause shear biases in Fourier\_Quad. A more careful study of this problem will be given in a future work. 

\subsection{Astrometric Calibration}
\label{astrometric_calibration}

Accurate astrometric calibration is crucial in weak lensing for at least two important reasons: 1. stacking of multi-exposure images; 2. calculation of the FDS, which should be subtracted from the shear estimator. The first point is not relevant in this paper, as our study here does not require stacking. The accuracy of the FDS distribution is our key concern. 
  
Although the formalism of \S\ref{fd} has been a routine in deriving the astrometric solution, we find that the fitting procedures sometimes do not converge well, leading to a failed astrometric solution. This happens in the public code SCAMP \citep{bertin2006} and THELI. This problem becomes more serious when the maximal order in the polynomial functions defined in eq.(\ref{mapping}) is large ($> 3$). This is unfortunate, as we need to increase the order of the polynomial functions to check the convergence/accuracy of the FDS distribution. For our purpose, we propose to modify the fitting procedure slightly: instead of performing the fitting on the projected plane $(\xi,\eta)$, we define another set of parameters $\mathrm{PU}$ to carry out the fitting on the re-scaled CCD plane $(\widetilde{X},\widetilde{Y})$ using the following formulae:
\begin{eqnarray}
\label{mapping2}
&&\widetilde{X}(\xi,\eta)=\xi+\mathrm{PU^1_3}r+\mathrm{PU^1_4}\xi^2+\mathrm{PU^1_5}\xi\eta+\mathrm{PU^1_6}\eta^2\\ \nonumber
&+&\mathrm{PU^1_7}\xi^3+\mathrm{PU^1_8}\xi^2\eta+\mathrm{PU^1_9}\xi\eta^2+\mathrm{PU^1_{10}}\eta^3+\mathrm{PU^1_{11}}r^3+... \\ \nonumber
&&\widetilde{Y}(\xi,\eta)=\eta+\mathrm{PU^2_3}r+\mathrm{PU^2_4}\eta^2+\mathrm{PU^2_5}\xi\eta+\mathrm{PU^2_6}\xi^2\\ \nonumber
&+&\mathrm{PU^2_7}\eta^3+\mathrm{PU^2_8}\xi\eta^2+\mathrm{PU^2_9}\xi^2\eta+\mathrm{PU^2_{10}}\xi^3+\mathrm{PU^2_{11}}r^3+... 
\end{eqnarray}
where $r=\sqrt{\xi^2+\eta^2}$. $(\widetilde{X},\widetilde{Y})$ can then be directly used to match the CCD position $(X,Y)$ defined in eq.(\ref{XYxy}) through a simple $\chi^2$ as:
\begin{eqnarray}
\label{chi2}
\chi^2&=&\sum_i\left[X(x_i,y_i)-\widetilde{X}(\xi_i,\eta_i)\right]^2/\sigma_X^2(i)\\ \nonumber
&+&\sum_i\left[Y(x_i,y_i)-\widetilde{Y}(\xi_i,\eta_i)\right]^2/\sigma_Y^2(i)
\end{eqnarray}
By minimizing this form of $\chi^2$, the values of ${\rm CD}^i_j$, ${\rm CRPIX}(1,2)$, and ${\rm PU}^i_j$ can be straightforwardly derived. Note that we have intentionally omitted the constant shifts and some linear terms in $\xi$ and $\eta$ in eq.(\ref{mapping2}) because they are degenerate with ${\rm CD}^i_j$ and ${\rm CRPIX}(1,2)$. This degeneracy is quite obvious in the definition of our $\chi^2$. The values of $(\xi_i,\eta_i)$ can be calculated from the RA and DEC of a source in the reference catalog, with the CRVAL values given by the header of the FITS file \citep{cg2002}. For simplicity in this work, $\sigma_{\rm X}(i)$ and $\sigma_{\rm Y}(i)$ are chosen to be the same constant values, and therefore omitted.

Another key ingredient in astrometric calibration is to match the sources in the reference catalog with those on the CCD. This is done before the fitting is carried out. We use the initial values of ${\rm CD}^i_j$, ${\rm CRPIX}(1,2)$ from the header of the fits file to transfer the source coordinates from the pixel space directly to the plane of $(\xi,\eta)$ (by setting ${\rm PU}^i_j=0$), and use the given values of ${\rm CRVAL}(1,2)$ to turn the source positions of the reference catalog into $(\xi,\eta)$ as well. By plotting the differences between the position vectors in these two groups, we can identify a concentrated area, indicating that a number of source pairs share similar difference vectors. These sources are used to refine the values of ${\rm CD}^i_j$ and ${\rm CRPIX}(1,2)$ (still keeping ${\rm PU}^i_j=0$). The above procedures are then repeated for improving and finalizing the identification of source pairs, which are ultimately used in the polynomial fitting for the astrometric solution. 

\begin{figure}
  \centering
  \includegraphics[width=9cm]{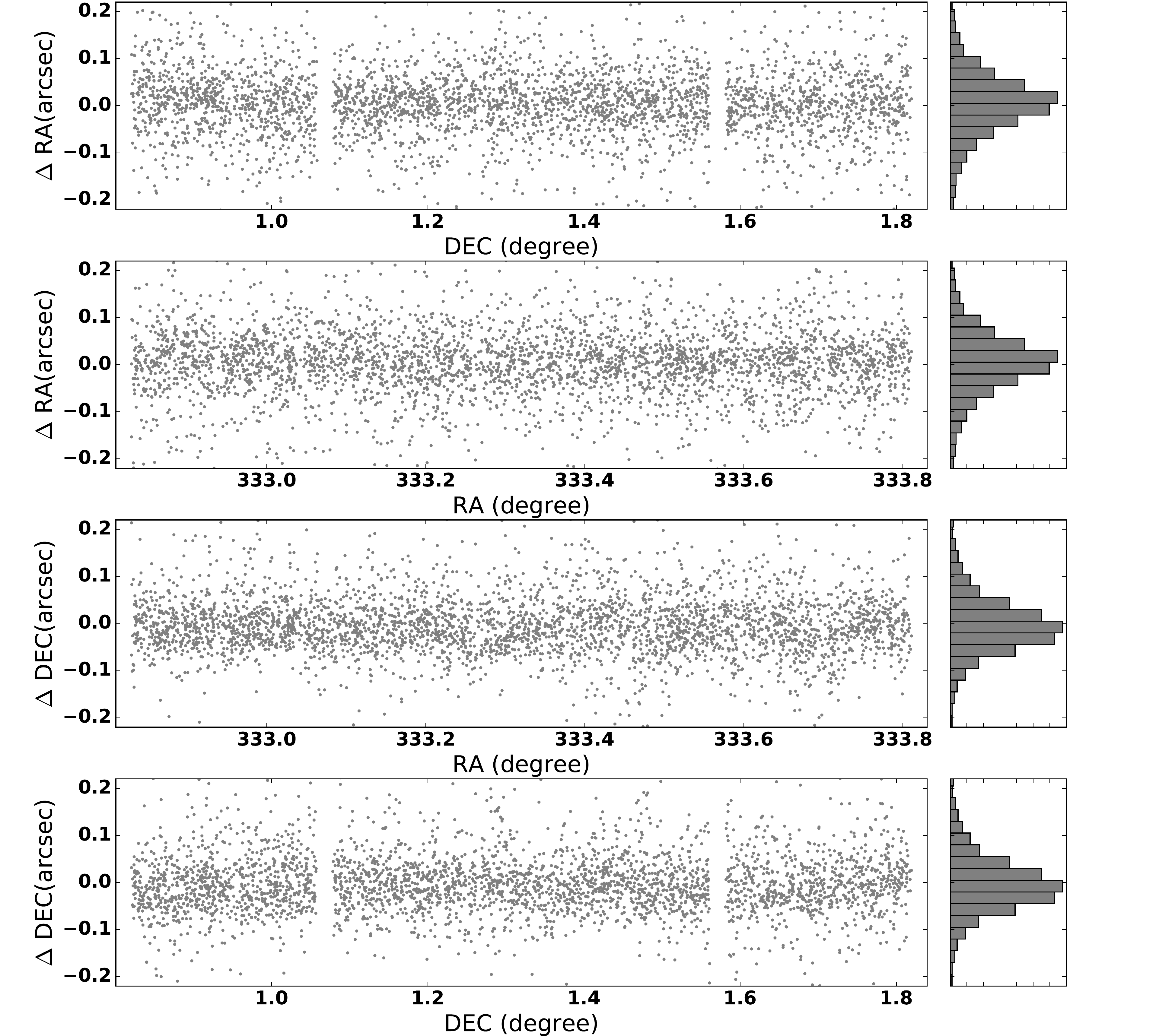} \\ \vspace{10pt}
  \caption{This is an example showing the performance of our astrometric calibration with the 3rd order polynomial fitting function for an exposure of the w4m0m0 field.}
  \label{fig:astrometry_order3}
\end{figure}

Fig.\ref{fig:astrometry_order3} shows the accuracy of the calibrated source positions for one exposure in the field of w4m0m0. The calibrated positions are compared to those listed in the reference catalog, which is chosen to be Gaia Dr2 \citep{gaia2018} in this study. We achieve the results using polynomial fitting functions up to the 3rd order, without involving terms proportional to $r$ or $r^3$. The rms values of the residuals in the four plots are about $0.05''$, and appear to be homogeneous over the whole sample. There are no visible systematic biases in the fitting.   

\begin{figure}
  \centering
  \includegraphics[width=9cm]{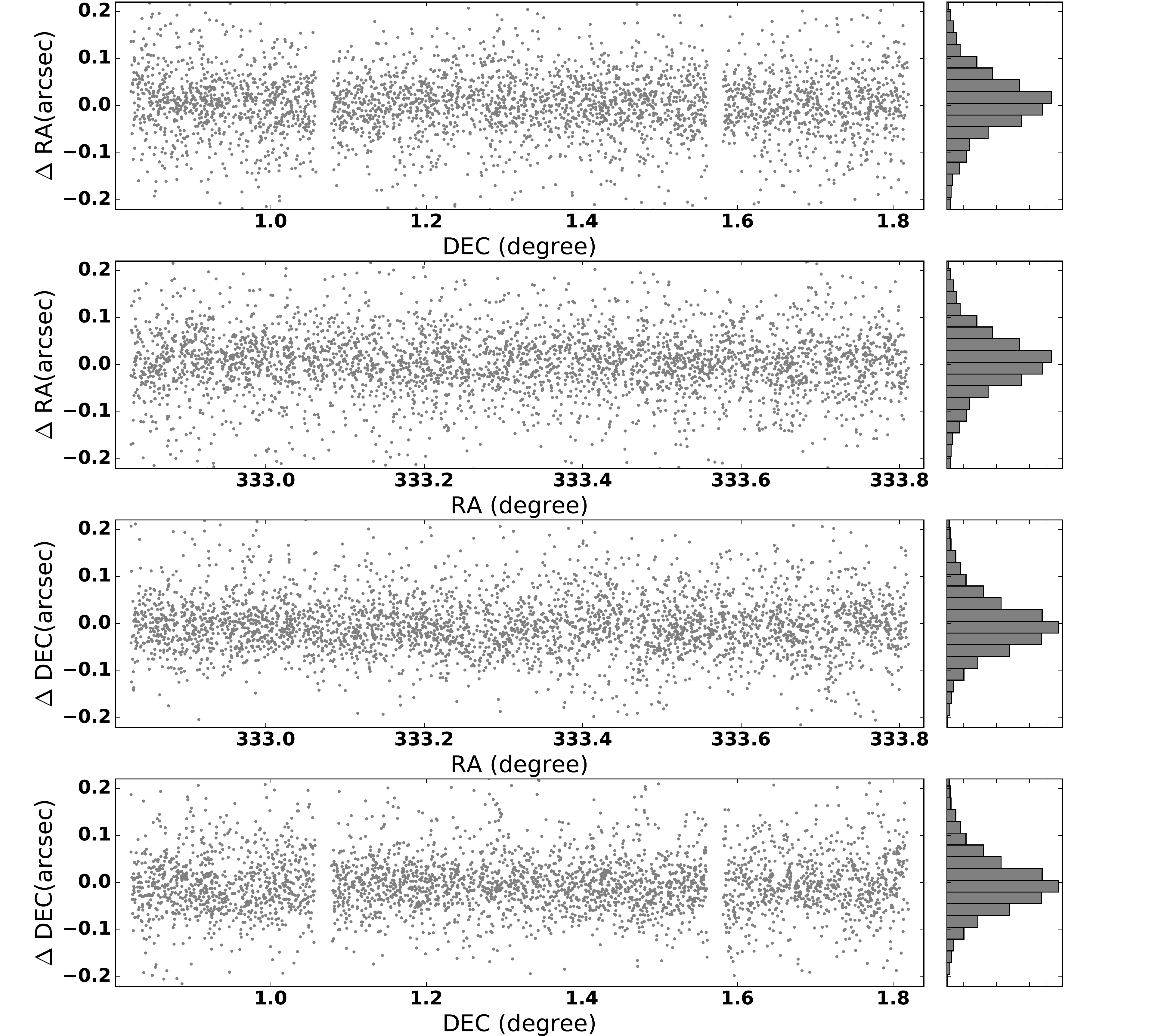}
  \caption{Same as fig.\ref{fig:astrometry_order3}, except that the polynomial fitting function is extended to the 9th order.}
  \label{fig:astrometry_order9}
\end{figure}

In fig.\ref{fig:astrometry_order9}, we show the results of the calibration for the same exposure using polynomial fitting functions up to the 9th order, still without the $r^{2n-1}$ terms ($n=1,2,3,4,5$). The overall quality of the fitting does not change much, implying that the astrometric calibration converges well at the 3rd order polynomial fitting in our pipeline. We also find that including the $r^{2n-1}$ terms does not make much difference. These facts ensure the accuracy of the FDS measurement for the purpose of this work.   

\subsection{Star Selection for PSF Reconstruction}
\label{star_finder}

As already demonstrated in \cite{lu2017}, for the CFHTLenS data, a chip-wise pixel-by-pixel spatial interpolation of the PSF power spectra with the 1st or 2nd order polynomial functions is the best way of reconstructing the PSF field for Fourier\_Quad. Our discussion here therefore only focuses on how to select out bright stars (typically with SNR $\gsim 100$) from bright sources.

Among the source stamps with large SNR, we realize that a significant fraction of them are indeed stars. It is also useful to note that the power spectra of stars have very similar profiles within a chip, and are more extended than those of galaxies. Using these facts, we adopt the following procedures to identify the stars: 

1. For each source stamp with SNR larger than $100$, we take its 2D power spectrum, and then apply on it the noise reduction, noise correction, and normalization procedures. The details of the noise reduction part are introduced in \S\ref{denoise}. Noise correction refers to the removal of the systematic impacts from the background noise and the Poisson noise on the source power spectrum according to the description in \S\ref{shear_measure}. The result of these procedures is to make the power spectrum image least affected by noise, and the power at $\mathbf{k}=0$ is unity;

2. A model PSF is constructed. Each of its pixel value is determined by sorting the corresponding pixel values from the power of all candidate sources, and taking the lower bound of the top 25\%;

3. The similarity between the images can be quantified by defining a distance $D$ between the normalized power spectra of two source stamps as:
\begin{equation}
\label{distance}
D_{nm}^2=\sum_{i=1}^{N}\left(I^n_i-I^m_i\right)^2k_i^{\alpha}
\end{equation}
where $I^{(n,m)}_i$ refers to the value of the $i^{th}$ pixel in the power image of the $n^{th}$ (or $m^{th}$) source stamp. $N$ is the total number of pixels used to evaluate $D$. $k_i^{\alpha}$ is simply a power law function of the wave number corresponding to the $i^{th}$ pixel, serving here as a weighting function. Since Fourier\_Quad uses the quadrupole moments of the power to estimate shear, we choose $\alpha=4$ to enhance the importance of pixels at large wave numbers. On the other hand, since the outskirts of the power image are dominated by noise, we limit the summation in eq.(\ref{distance}) to be within the central $25\times 25$ region;

4. Calculate the distance $D_{iM}$ between the model PSF and the $i^{th}$ source power. We then sort the values of $D_{iM}$, and take the upper bound of the bottom 25\% to be the threshold $D_c$. Sources with $D_{iM}>3D_c$ are removed from the candidate list, as they are simply too different from the model PSF;

5. With the remaining candidate sources, we perform the pixel-by-pixel 2nd-order polynomial fitting over the chip scale to construct the model PSF that varies with position. $D_{iM}$ is then newly defined as the distance between the $i^{th}$ source and the model PSF at its position. We define $\sigma$ as $\sqrt{\sum_iD_{iM}^2/N_T}$, \ie, the rms of the distances, with $N_T$ being the number of candidates. Sources with $D_{iM}$ greater than $2\sigma$ are removed from the candidate list;

6. Step 5 is repeated until the candidate list is not changed anymore, or the number of sources in the candidate pool is below a certain threshold. The threshold must at least allow us to perform a valid 2nd-order polynomial fitting. Note that if step 5 is stopped due to the second reason, there are simply no PSF or shear measurements further carried out on that chip. 

\begin{figure}
  \centering
  \includegraphics[width=8cm]{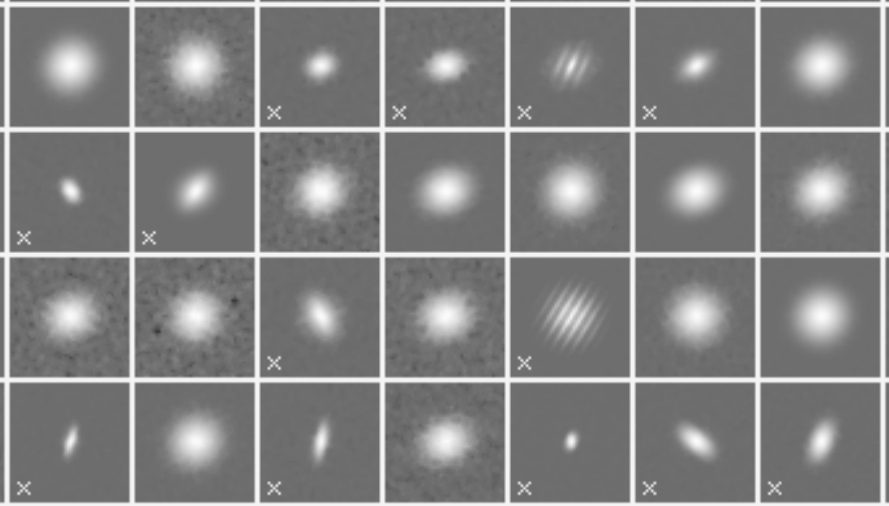}\vspace{0.1in}
  \includegraphics[width=8cm]{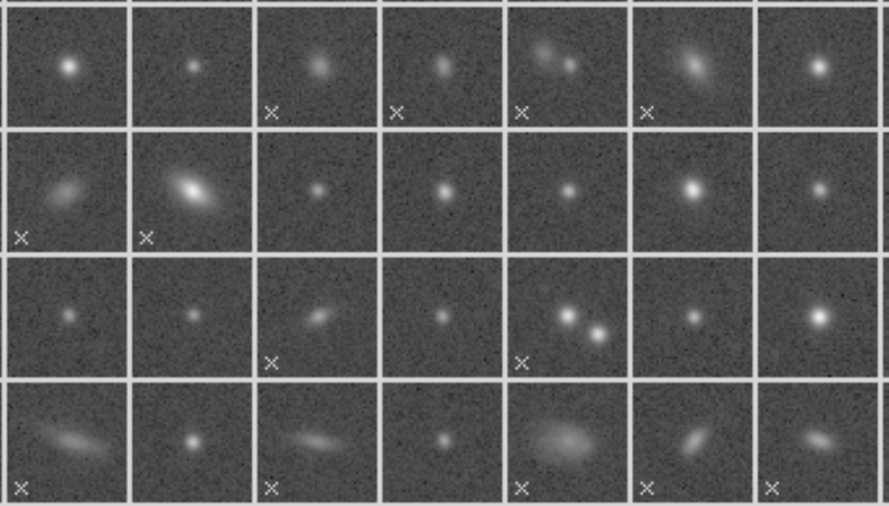}
  \caption{As an example, we show source images (lower panel) and their corresponding normalized power spectra (upper panel) that are collected as candidates for stars in the field of w2p2p1. Those with the cross-marks on the lower-left corners are designated as nonstellar objects according to the algorithm of \S\ref{star_finder}. }
  \label{star_selection}
\end{figure}

Fig.(\ref{star_selection}) shows examples of source images (lower panel) and their corresponding power spectra (upper panel) that are selected as star candidates on a chip of the w2p2p1 field. Those without the cross-marks on the lower-left corners are identified as stars. The example shows that our algorithm is quite accurate in picking out the right sources (stars) for PSF reconstruction. It is evident from the figure that the nonstellar objects are more extended in real space, and more compact in Fourier space. We have checked a large number of star images by eye, and have not found significant problems. 

\subsection{Noise Reduction}
\label{denoise} 

There are various ways of reducing the image noise, though not all of them are suitable for accurate shear measurement. For the method of Fourier\_Quad, we find that simple polynomial fittings in the Fourier space perform reasonably well in reducing the noise for the CFHTLenS data. Its accuracy (especially on the faint objects), though, is still subject to more numerical tests in our ongoing work. Our de-noising operations are applied on the power spectrum of the image. Each pixel is re-evaluated by fitting its neighboring $5 \times 5 $ region with a 2nd order polynomial function. We repeat this fitting for every pixel. For pixels close to a boundary, the fitting region also include pixels from the opposite side due to periodicity. For better results, we adopt the following three tricks: 1. rather than fitting the pixel values, we fit their logarithms, which generally form a much smoother function; 2. we exclude the four vertices of the fitting region to reduce its anisotropy; 3. the pixel of $k_x=k_y=0$ is not included in any fitting, for its value is strongly affected by the residual background level.   

\begin{figure}
  \centering
  \includegraphics[width=8cm]{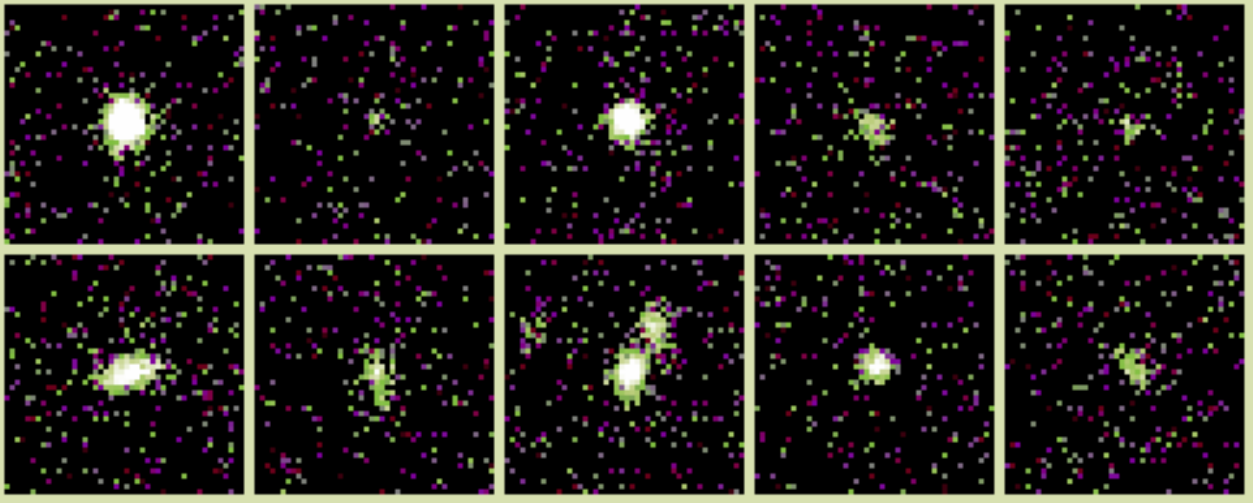}\vspace{0.1in}
  \includegraphics[width=8cm]{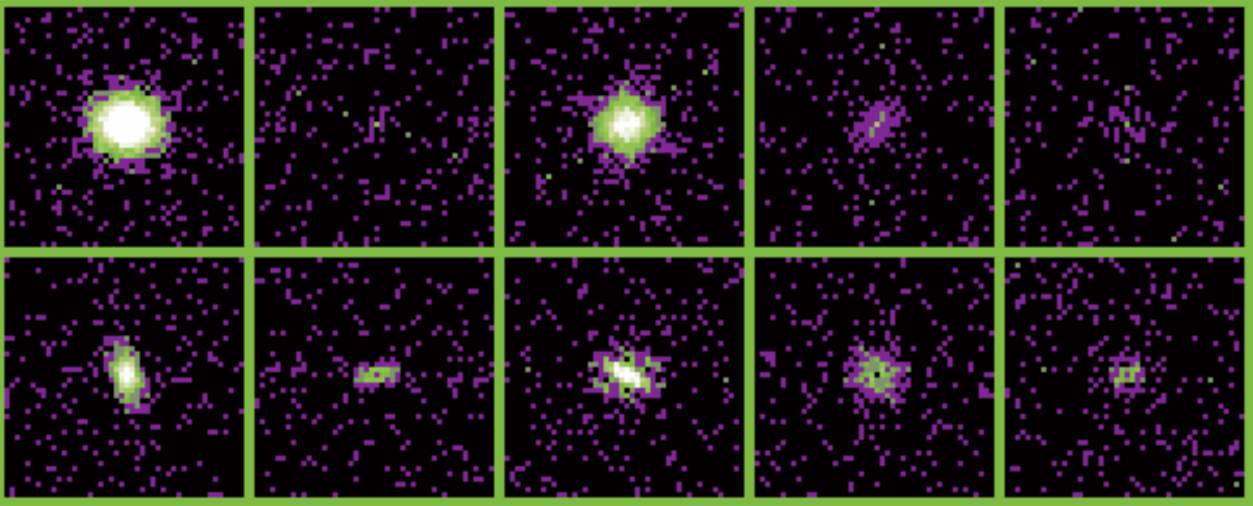}\vspace{0.1in}
  \includegraphics[width=8cm]{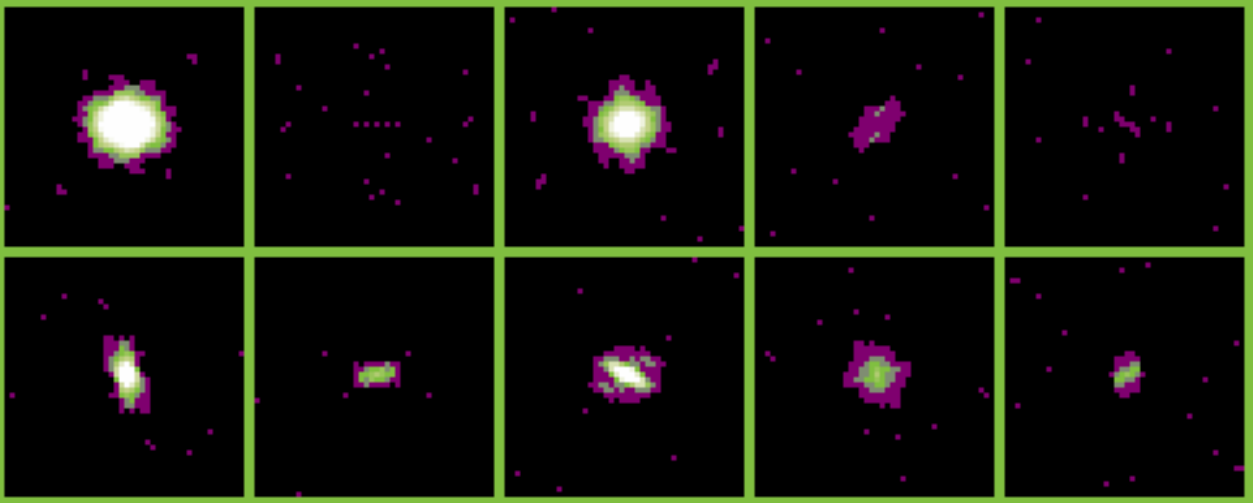}
  \caption{Examples showing the performance of our noise reduction procedure introduced in \S\ref{denoise}. The upper panel shows the original source images. The middle and lower panels present their corresponding unsmoothed and smoothed power spectra respectively.}
  \label{noise_reduction}
\end{figure}

Several examples are shown in fig.\ref{noise_reduction}, in which the upper panel shows the original source images, and the middle and lower panels present their corresponding unsmoothed and smoothed power spectra.
    
\section{Results}
\label{results}

Our main results are presented with two different source catalogs: one is generated by our own pipeline, named as ``FQ'' hereafter (representing Fourier\_Quad); another is downloaded from the website of the CFHTLenS project, named as ``LF'' (standing for LensFit). Source selection in the current version of our pipeline is based on individual exposures with the procedures described in \S\ref{source_select}. In the second case, from the LF catalog, we take the source position (RA, DEC) to find the corresponding CCD position (X, Y) in each exposure using the astrometric parameters. The validities and boundaries of the sources are then determined with the procedures of \S\ref{source_select} as well. Note that we do not use the galaxy ellipticities and weights in the LF catalog.
 
It is worth noting that in Fourier\_Quad, it is not necessary to perform a comprehensive star-galaxy separation, as the inclusion of point sources in the galaxy sample does not affect the results of either shear stacking or shear-shear correlations, as shown in \cite{zzl2017}. This is a unique and useful feature of Fourier\_Quad. It is also supported by the results of this work, as demonstrated in this section.

\begin{figure}
  \centering
  \includegraphics[width=8cm]{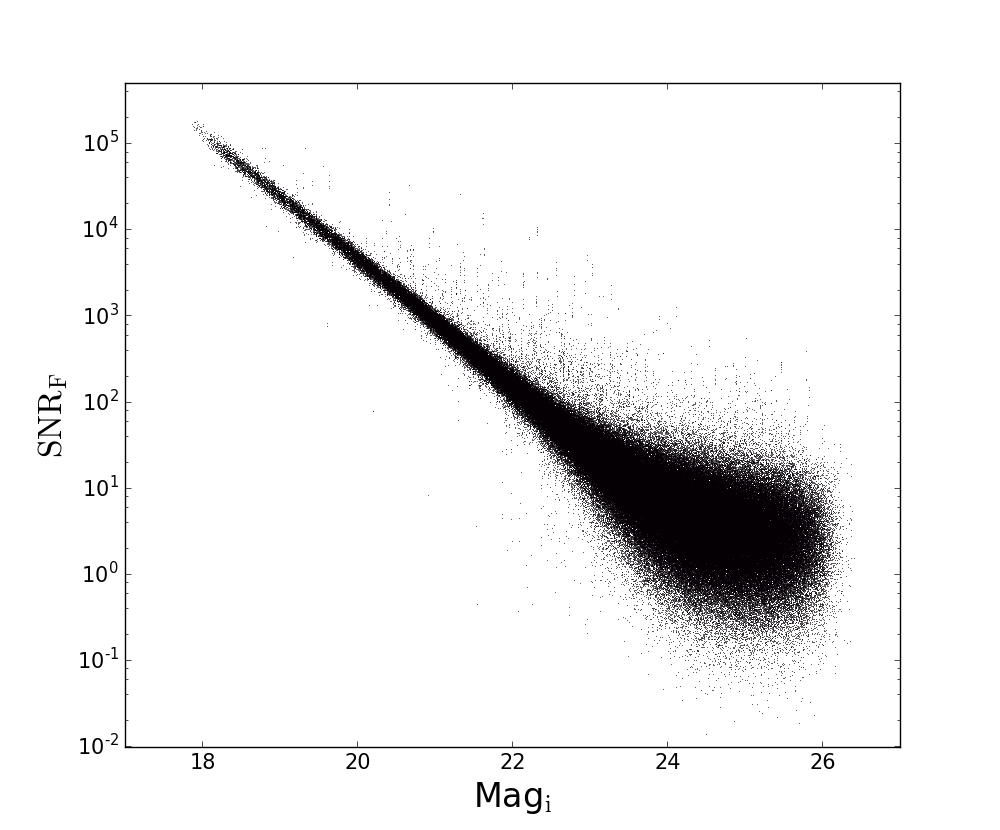}
  \caption{The relation between $\mathrm{SNR_F}$ and the i-band magnitude using the data from the field of w2m0m0.}
  \label{SNR_MAG}
\end{figure}

Finally, we are aware that an inappropriate selection of the galaxy sample may lead to shear biases that are purely due to the selection rule itself. This is caused by the correlation between the selection function and the shear signal, and typically called the "selection effect". Previous tests of the Fourier\_Quad method are based on simulated images with intrinsically random morphologies, which are by definition free from the selection effect. For real data, selections are inevitable at both the bright/large and faint/small ends. The latter is especially important, as the faint end involves a large fraction of the total galaxy population. Our idea is to use the total flux of the source as a selection function, as it is least affected by lensing/shear. The actual quantity we use to represent flux is $P_0$ (indeed flux$^2$), \ie, the power of the source image at $\mathbf{k}=0$. Note that the noise in the source power spectrum is reduced by the procedures defined in \S\ref{denoise}. To make $P_0$ dimensionless, we divide it by the average power $\bar{P}_N$ at large wave numbers (estimated at the boundary of the Fourier domain), \ie, the power of the Poisson noise. Our selection function can be expressed as a Fourier-domain signal-to-noise-ratio $\mathrm{SNR_F}=P_0/\bar{P}_N$. A detailed discussion of $\mathrm{SNR_F}=P_0/\bar{P}_N$ as well as other selection functions will be given in another work (Li et al., in preparation). The relation between $\mathrm{SNR_F}$ and the i-band magnitude of the sources are given in fig.\ref{SNR_MAG}, with data from the field of w2m0m0. The figure shows that $\mathrm{SNR_F}$ has a very good monotonic relation with magnitude at the bright end. The scatter of the relation becomes much more significant at the faint end, implying that a cut in $\mathrm{SNR_F}$ can be very different from a cut in magnitude for source selection.

\subsection{The FQ Catalog}

\begin{figure}
  \centering
  \includegraphics[width=8cm]{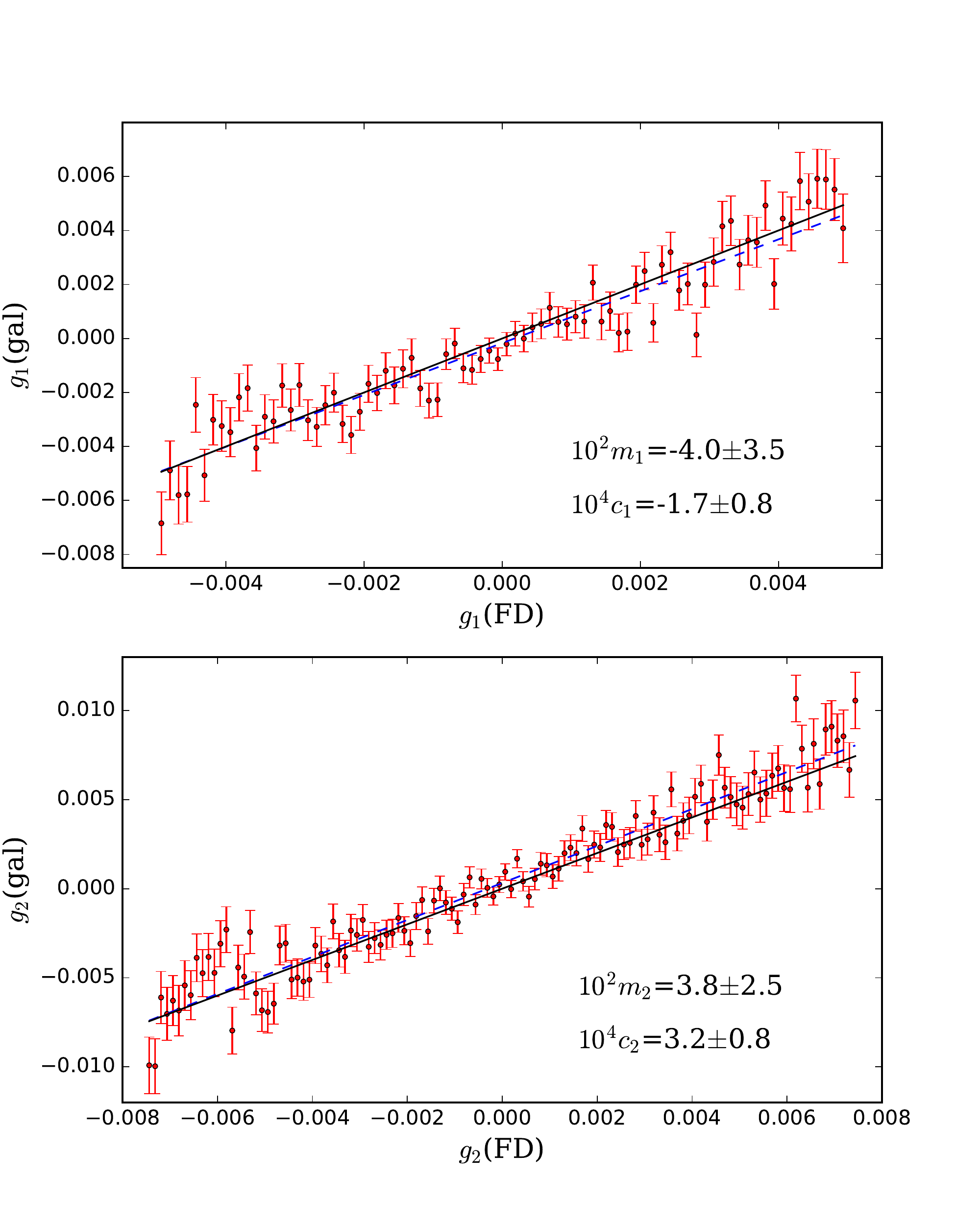}
  \caption{Comparison of the FDS with the shear signals recovered with galaxy shapes by the Fourier\_Quad method. The sources are from the FQ catalog. The black lines are the ``$y=x$'' function, and the red data points with $1\sigma$ error bars are the recovered shear values from the source images of CFHTLenS, best-fitted by the blue-dashed lines. The best-fitting multiplicative and additive bias parameters are listed at the lower-right corner of each panel.}
  \label{result_FQ}
\end{figure}

Fig.\ref{result_FQ} shows how well the FDS can be recovered with sources from the FQ catalog. The black lines are simply the ``$y=x$'' function, and the data points with $1\sigma$ error bars are the recovered shear values from source images of CFHTLenS, best fitted by the blue-dashed lines. The best-fitting parameters $(m,c)$ are defined as: $g_{1,2}(\mathrm{gal})=g_{1,2}(\mathrm{FD})*(1+m_{1,2})+c_{1,2}$, and are listed on the lower-right corner of each panel. The sources are binned according to the FDS. Note that the ranges of $g_1(\mathrm{FD})$ and $g_2(\mathrm{FD})$ are both very small, \ie, [-0.005,0.005] and [-0.0075, 0.0075] respectively, quite suitable for testing the accuracy of shear recovery at the level of cosmic shear. We divide these ranges into about 100 bins for each shear component to take advantage of the large source number available from CFHTLenS. In this plot, we use sources with $\mathrm{SNR_F}\geq 4$, and the number of bright stars for PSF reconstruction on the chip of the source is required to be larger than 20. There are no additional cuts applied on the source sample, and all sources that pass the validity check of \S\ref{source_select} have valid measurements of the $G_{1,2}, N, U, V$ shear parameters defined in eq.(\ref{shear_estimator},\ref{def_U}). There are $42$ million sources in total. We use the PDF-SYM method to derive the shear signals in the figure. Note that images of the same galaxy on different exposures are treated as different sources in this work. The results show that the multiplicative biases are at the level of $\lsim 4\%$ [$m_1=(-4.0\pm3.5)\times 10^{-2}$, $m_2=(3.8\pm 2.5)\times 10^{-2}$]. We caution that the data points seem to have slight systematic deviations from the ``$y=x$'' relation at large values of $g_2(\mathrm{FD})$ (close to the four corners of an exposure), which drive $c_2$ somewhat larger than we expected [$c_2=(3.2\pm 0.8)\times 10^{-4}$, $4\sigma$ significance]. This problem deserves further investigation in a future work.

\begin{figure}
  \centering
  \includegraphics[width=8cm]{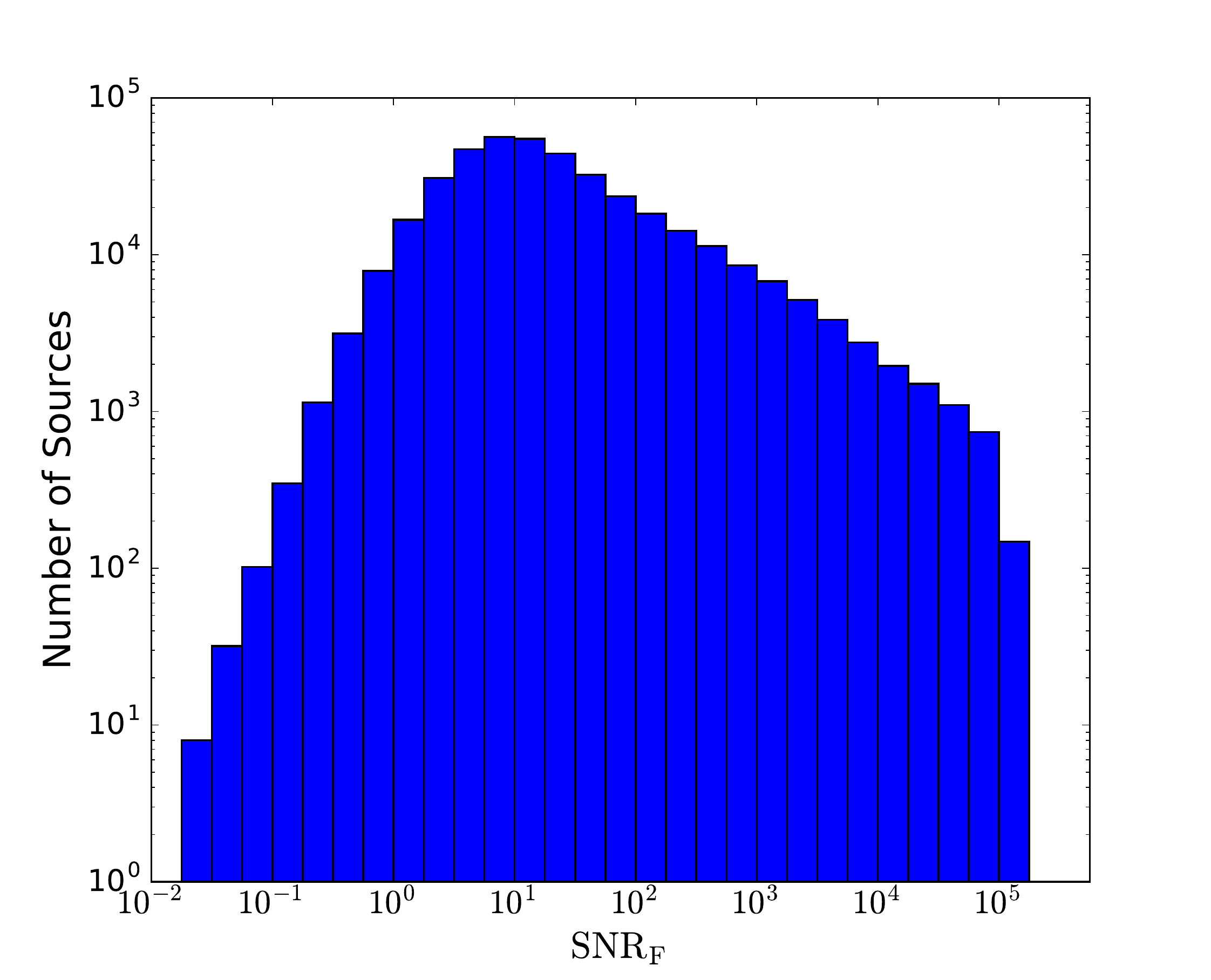}
  \caption{The distribution of $\mathrm{SNR_F}$ in the field of w2m0m0. The sources are from the FQ catalog.}
  \label{SNR_F_dist}
\end{figure}

We choose $\mathrm{SNR_F}\geq 4$ as our selection criteria because sources with much smaller $\mathrm{SNR_F}$ seem to be strongly affected by the source-locating (or pre-selection) procedures defined in \S\ref{source_select}. This can be seen, \eg, in fig.\ref{SNR_F_dist} from the field of w2m0m0. 

\subsection{The LF Catalog}

\begin{figure}
  \centering
  \includegraphics[width=8cm]{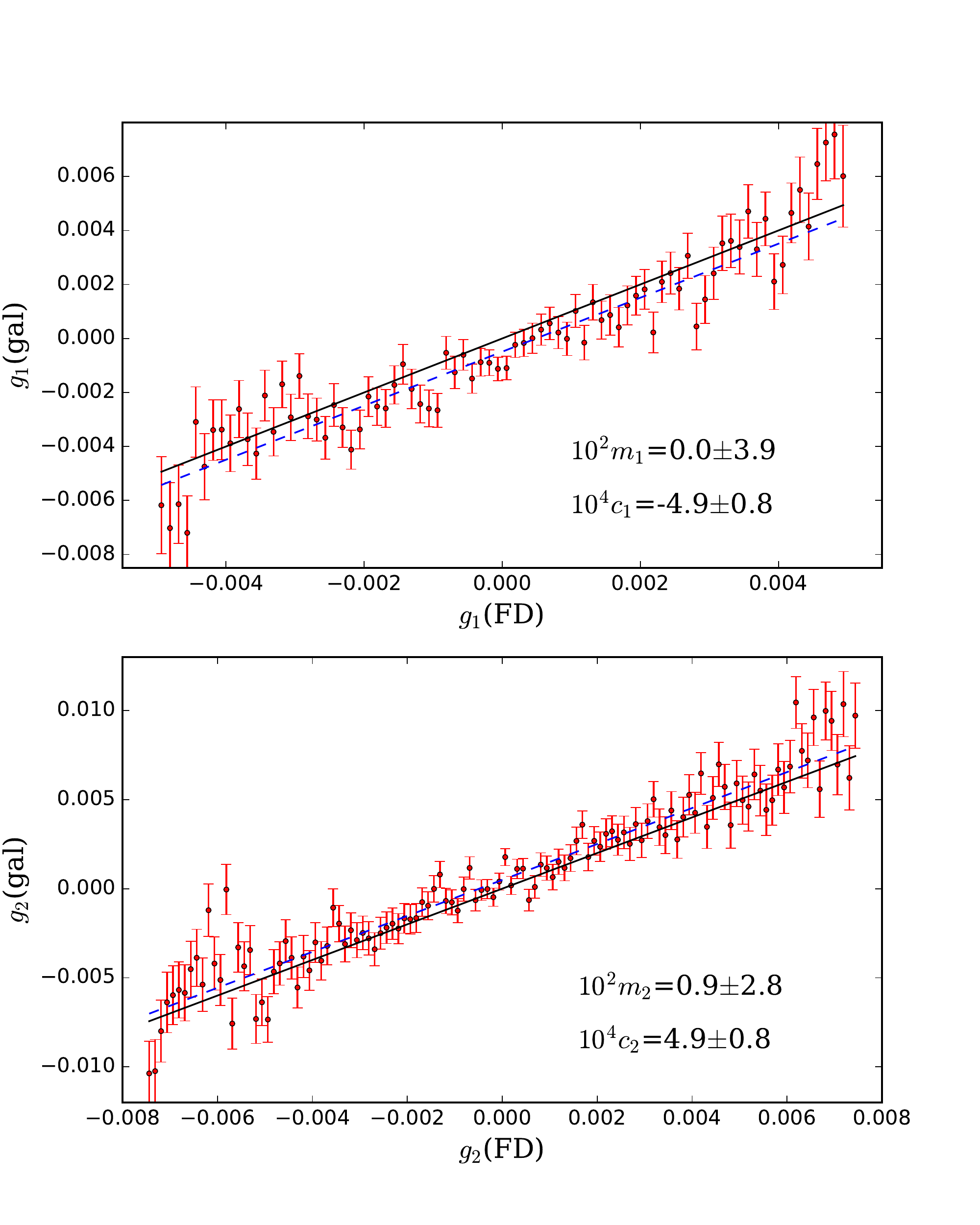}
  \caption{Similar to fig.\ref{result_FQ}, but with source positions given by the LF catalog.}
  \label{result_LF}
\end{figure}

Fig.\ref{result_LF} is similar to fig.(\ref{result_FQ}), but with sources defined in the LF catalog. We apply the same cut on the $\mathrm{SNR_F}$ value, and the same requirement on the star number of the chip for PSF reconstruction. In total, we get $39$ million source images. The results of fig.\ref{result_LF} look reasonably good. It shows insignificant multiplicative biases, but slight additive biases [$c_1=(-4.9\pm 0.8)\times 10^{-4}$, $c_2=(4.9\pm 0.8)\times 10^{-4}$]. It is not yet clear what causes the additive biases. It is possibly due to the differences in the source identification parts of the two pipelines: sources in the LF catalog are identified on co-added images, while Fourier\_Quad relies on exposures independently. This issue will be studied more carefully in another work. For Fourier\_Quad, source-identification based on multiple exposures are still under development. 

\begin{figure}
  \centering
  \includegraphics[width=8cm]{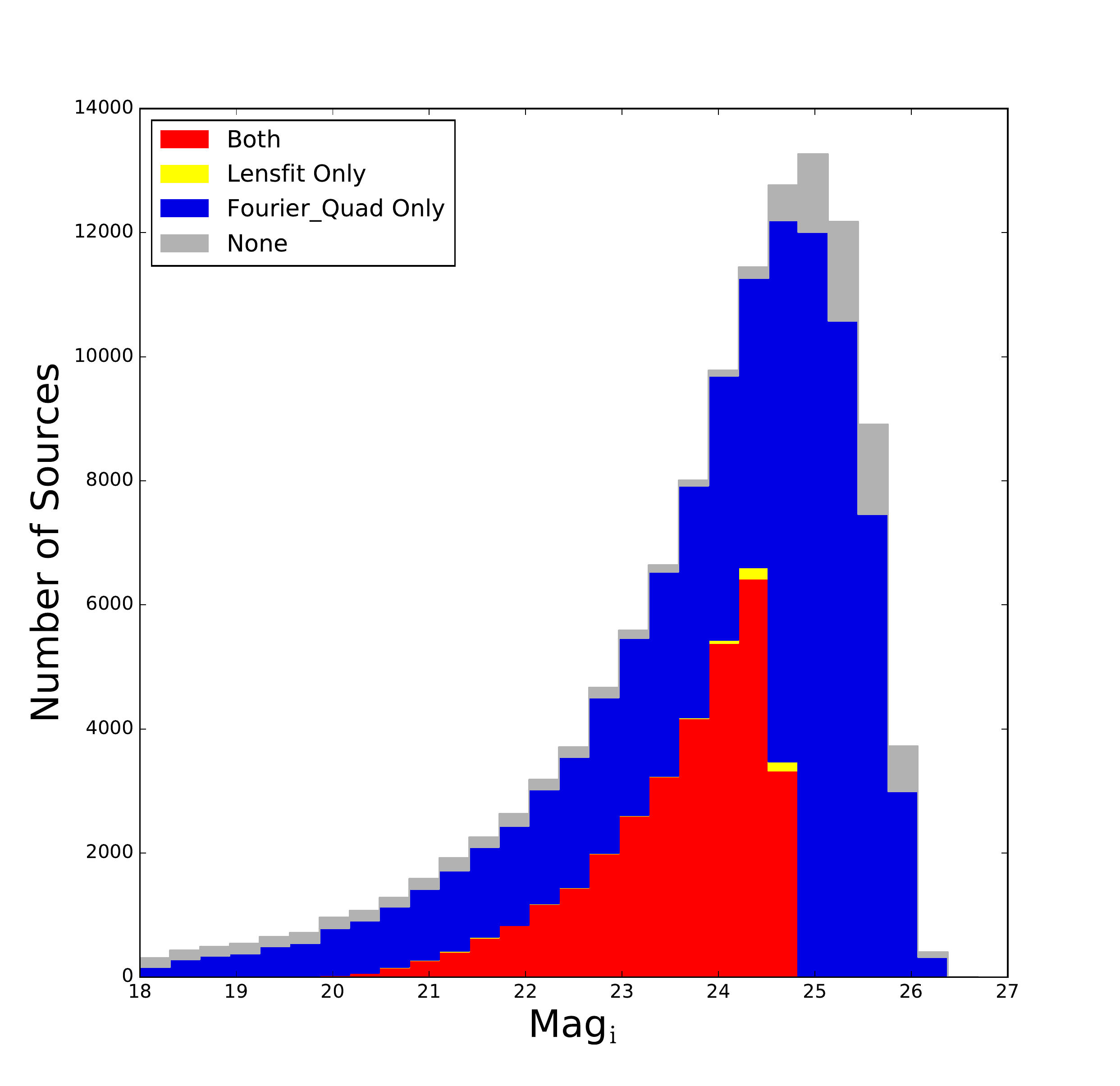}
  \caption{The i-band magnitude distributions of sources with different shear measurement status. The red population has valid shear measurements from both Lensfit and Fourier\_Quad; the yellow and blue ones only have Lensfit and Fourier\_Quad contributions respectively; the gray ones have no shear measurements at all. The data is from the field of w2m0m0.}
  \label{comp_w2m0m0}
\end{figure}

Fig.\ref{comp_w2m0m0} shows the i-band magnitude distribution of sources with different shear measurement status. The red ones have valid measurement from both Lensfit and Fourier\_Quad, and the yellow and blue ones only have contributions from Lensfit and Fourier\_Quad respectively. The gray population does not have shear measurements at all. The data is from the field of w2m0m0. It is encouraging to note that our Fourier\_Quad pipeline can provide valid shear measurements on a much larger galaxy population than that by Lensfit. 

\begin{figure}
  \centering
  \includegraphics[width=8cm]{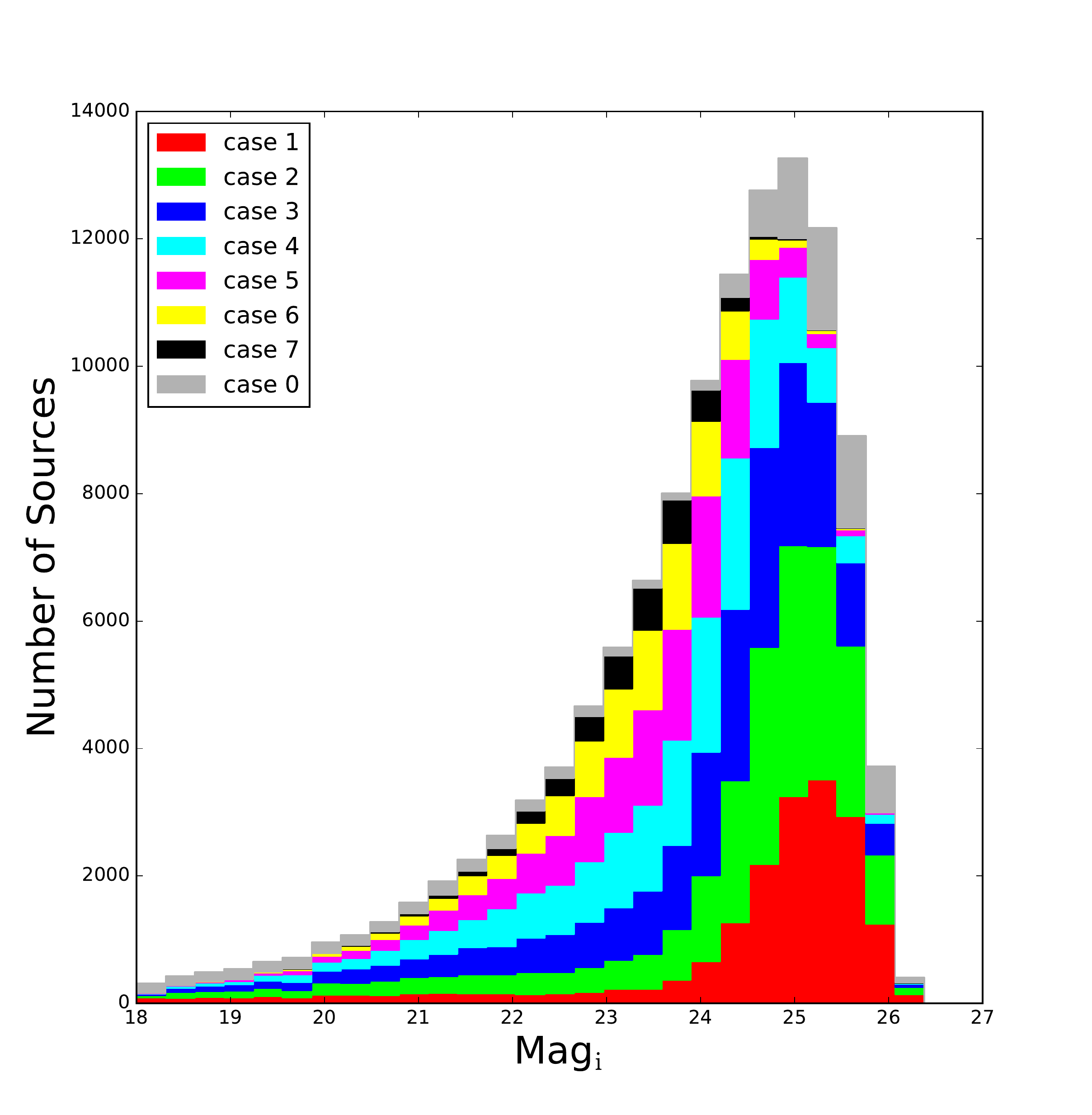}
  \caption{The i-band magnitude distributions of sources with different shear measurement status in Fourier\_Quad. Case "n" means there are "n" valid shear measurements/exposures associated with a source. The data is from the field of w2m0m0.}
  \label{w2m0m0_expos}
\end{figure}

For each source, Fourier\_Quad makes an independent shear measurement on every distinct exposure as long as there are no image defects at the source position. Fig.\ref{w2m0m0_expos} shows the distribution of the number of valid exposures for sources of different magnitudes. In the figure, ``case n'' stands for ``n'' valid exposures, represented by different colors in the histogram. It is not surprising that at the faint end, a source can be identified only on very few exposures.  

\section{Conclusion and Discussion}
\label{conclusion}

We have presented a way of testing shear recovery accuracy on real images. The target "shear signals" are given by the field distortion effect, which is naturally involved in any optical system. The distribution of the FDS can be accurately mapped out on the CCD plane with the astrometric parameters. These shear signals are recoverable with galaxy images whenever there are a large number of exposures available. The cosmological shear signals in this case are generally not relevant. By comparing the FDS with the stacked galaxy shear estimators, one can directly quantify the multiplicative and additive shear biases on real images.

As an example, we show the performance of the Fourier\_Quad method using the CFHTLenS data. The FDS values on MegaCam are around $0.005$ or less, very suitable for shear calibration at the level of cosmic shear. We have described in \S\ref{reduction} a number of important details in the image processing pipeline of Fourier\_Quad, including background removal, source identification and deblending, astrometric calibration, star selection for PSF reconstruction, noise reduction. Our pipeline can also read in source locations (RA \& DEC) from an external source catalog. 

Our main results are presented in fig.\ref{result_FQ} and fig.\ref{result_LF} using two different source catalogs: the FQ one is by our own pipeline, and the LF catalog is from the official data release of CFHTLenS. The shear signals in both cases are measured by Fourier\_Quad. Overall, the multiplicative biases are at the level of $\lsim 4\%$. The results from the sources of the LF catalog reveal some slight additive biases of about $5\times10^{-4}$ ($6\sigma$ significance) for both shear components. This problem is minor for the FQ catalog: only $g_2$ has an additive bias of $3.2\times 10^{-4}$ ($4\sigma$). We believe the difference is due to some subtleties in the source locating/identification part of the pipeline of CFHTLenS, which is very different from ours. The most significant difference is that sources in the LF catalog are identified on co-added images, while our pipeline so far processes every exposure individually. A combinatory multi-exposure version of Fourier\_Quad is still under development.

Within the LF catalog provided by CFHTLenS, we find that there are a large portion of sources that have valid magnitudes, but not shear measurements. This problem is partially remedied by our Fourier\_Quad pipeline, which provides valid shear measurements for most of the sources in the LF catalog, as shown in fig.\ref{comp_w2m0m0}. In principle, for each source, Fourier\_Quad can yield one shear measurement from each exposure. The reality is that some sources in the LF catalog cannot be properly identified on single exposures in Fourier\_Quad, especially at the faint end as shown in fig.\ref{w2m0m0_expos}. This is due to their weak significance or image defects. 

For valid source images, their shear estimators can all be measured by Fourier\_Quad. The only exception is when there are not enough bright stars on the chip for PSF reconstruction. The shear recovery part of our pipeline simply does not apply any selection rules on the source morphology (shape, size, etc.), allowing us to avoid complicated decisions regarding selection effects due to shear measurement itself. Note that it is not even necessary to exclude stellar objects or point sources from the galaxy samples in Fourier\_Quad. We consider this a significant advantage of our pipeline. 

Our proposal of shear testing with field distortion should be useful for any shear measurement algorithm, including some recently developed ones \citep{sh2017,hm2017,of2017,tewes2018,pujol2018,li2018}. It is better to work with individual exposures, in which case the field distortion signals are well defined. For algorithms like Lensfit, which simultaneously fits the source shape on multiple exposures, the field-distortion signals are already included in the forward modeling of the galaxy image on each exposure. In this case, the recovered ellipticities can still be plotted against the average field-distortion signal to form a null test, assuming the spatial offsets of the relevant exposures on the sky are much smaller than their sizes. Further discussion of these topics is beyond the scope of this work.

It is also important to note that accurate astrometric calibration is indispensable for a successful shear measurement program. We have introduced a modification to the fitting between the projected sky plane and the CCD plane in \S\ref{astrometric_calibration}, enabling us to easily extend the fitting functions to high order polynomials, and to check the convergence of the astrometric solution. A more comprehensive comparison between the astrometry part of our pipeline and the standard software, such as SCAMP, will be provided in a separate work. We also caution that polynomial fitting is not likely good enough for recovering distortions caused by instrumental effects at special locations \citep{bernstein2017} of the CCD. For the CFHTLenS data, we have not studied such a problem carefully. We leave a detailed discussion of this topic to a future work.

Finally, it is encouraging to note that our Fourier\_Quad pipeline is now able to carry out shear measurement from almost raw CCD images (after flat-field correction). The overall image processing speed is very fast. For example, for the CFHTLenS data, overall, it only takes about 0.02 CPU*seconds for each galaxy image on average. We therefore consider the Fourier\_Quad pipeline a very promising tool for probing the cosmic structure in the ongoing and planned large scale galaxy surveys. In the future development of our pipeline, we will try to include the following components: 1. source identification using information from multiple exposures; 2. ways of stacking under-sampled images for accurate shear recovery (mostly for space-based missions); 3. treatments of instrumental effects \citep{rhodes2010,antilogus2014}.

\acknowledgments{This work is based on observations obtained with MegaPrime/MegaCam, a joint project of CFHT and CEA/DAPNIA, at the Canada-France-Hawaii Telescope (CFHT) which is operated by the National Research Council (NRC) of Canada, the Institut National des Sciences de l\'Univers of the Centre National de la Recherche Scientifique (CNRS) of France, and the University of Hawaii. This research used the facilities of the Canadian Astronomy Data Centre operated by the National Research Council of Canada with the support of the Canadian Space Agency. CFHTLenS data processing was made possible thanks to significant computing support from the NSERC Research Tools and Instruments grant program.

In the early stage of this project, the processing of single exposure images is conducted using the THELI software, a tool for the automated reduction of astronomical images developed by the CFHTLenS team \citep{erben2005,schirmer2013}.

JZ is supported by the National Key Basic Research and Development Program of China (No.2018YFA0404504), the NSFC grants (11673016, 11433001, 11621303), and the National Key Basic Research Program of China (2015CB857001). Xiangchong Li is supported by Global Science Graduate Course (GSGC) program of University of Tokyo. LPF acknowledges the support from NSFC grants 11673018, 11722326 \& 11333001; STCSM grant 16ZR1424800 \& 188014066; and SHNU grant DYL201603. D.Z.Liu and Z.H.Fan are supported in part by NSFC of China under the grants 11333001 and 11653001. G.L. is supported by NSFC (No.11673065,11273061 and 1133008) and the 973 program  (No. 2015CB857003).

}

\end{document}